\documentclass[journal]{IEEEtran}
\usepackage{amsmath,amsfonts}
\usepackage{algorithmic}
\usepackage{algorithm}
\usepackage{array}
\usepackage[caption=false,font=normalsize,labelfont=sf,textfont=sf]{subfig}
\usepackage{textcomp}
\usepackage{stfloats}
\usepackage{url}
\usepackage{verbatim}
\usepackage{graphicx}
\usepackage{cite}

\usepackage{siunitx}
\usepackage{booktabs}
\usepackage{multirow}
\usepackage{graphicx}
\usepackage{soul, color, xcolor}

\begin{document}

\title{A Novel Semisupervised Contrastive Regression Framework for Forest Inventory Mapping with Multisensor Satellite Data}

\author{Shaojia~Ge,~Hong~Gu,~Weimin~Su,~Anne~L\"onnqvist,~Oleg~Antropov
\thanks{This work was supported by the National Natural Science Foundation of China under Grant 62001229, Grant 62101264 and Grant 62101260.}
\thanks{Shaojia Ge, Hong Gu, and Weimin Su were with Department of Electronic Engineering, School of Electronic and Optical Engineering, Nanjing University of Science and Technology, 210094 Nanjing, China, email: geshaojia@njust.edu.cn, guhong666@njust.edu.cn, suweimin@njust.edu.cn.}
\thanks{{Anne L\"onnqvist} and Oleg Antropov were with VTT Technical Research Centre of Finland, Espoo, Finland, email: name.surname@vtt.fi.}
}

\markboth{Journal of \LaTeX\ Class Files,~Vol.~, No.~, November~2022}%
{Shell \MakeLowercase{\textit{et al.}}: A Sample Article Using IEEEtran.cls for IEEE Journals}


\maketitle

\begin{abstract}
Accurate mapping of forests is critical for forest management and carbon stocks monitoring. Deep learning is becoming more popular in Earth Observation (EO), however, the availability of reference data limits its potential in wide-area forest mapping.
To overcome those limitations, here we introduce contrastive regression into EO based forest mapping and develop a novel semisupervised regression framework for wall-to-wall mapping of continuous forest variables. It combines supervised contrastive regression loss and semi-supervised Cross-Pseudo Regression loss. The framework is demonstrated over a boreal forest site using Copernicus Sentinel-1 and Sentinel-2 imagery for mapping forest tree height. Achieved prediction accuracies are strongly better compared to using vanilla UNet or traditional regression models, with relative RMSE of 15.1\% on stand level. We expect that developed framework can be used for modeling other forest variables and EO datasets.  

\end{abstract}

\begin{IEEEkeywords}
deep learning, contrastive regression, image time series, regression, Sentinel-1, Sentinel-2, boreal forest, tree height
\end{IEEEkeywords}

\section{Introduction}
\IEEEPARstart{R}{ecently}, deep learning (DL) has gained considerable attention in environmental remote sensing \cite{zhu2020}, particularly in forest mapping applications \cite{lang2019country, ge2022improved, ge2022lstm}. In model based forest mapping, the key focus is on modeling the relationship between measured reference and EO data \cite{mcrob2007}. While DL models normally require large amounts of reference data, measured forest reference data are often insufficient.
On the other hand, volume of available EO data samples largely exceeds amount of reference forest labels, and relationships between EO data measurements not covered by reference labels are typically ignored. These issues limit the potential of DL models in forest mapping. 
   
Self-supervised learning (SSL), especially contrastive learning (CL), is particularly suitable to overcome such reference data limitations \cite{liu2021self}, by successfully learning representative features between unlabeled samples \cite{chen2020simple, he2020momentum}.
Furthermore, CL was also demonstrated in supervised scenarios with improved classification accuracy \cite{khosla2020supervised}.
This motivates further use of CL in various remote sensing applications and some considerable performance improvements have been reported \cite{wang2022self}.
However, all the reported studies focused on semantic segmentation or change detection tasks.
Studies on the use of CL in modeling relationships between continuous environmental variables and EO data are missing, particularly in mapping forest variables using multisensor EO data. 

To bridge this gap, in this letter we introduce a novel contrastive regression loss (CtRL) \cite{wang2022contrastive} suitable for regression tasks such as forest mapping.   
Different from InfoNCE loss \cite{oord2018representation}, which is usually employed in classification situations, CtRL is based on the assumption that intrinsic relationship between labels can be useful to reveal the proximity between normalized features on embedding hypersphere \cite{khosla2020supervised, wang2020hypersphere}.
With the help of CtRL, relative correlations between different samples can be further exploited. 
Further, a hybrid framework combining both supervised CtRL and semi-supervised Cross-Pseudo Regression loss (CPR) is established.
Three kinds of input features are leveraged in this framework: a) supervised semantic features brought by reference data, b) unsupervised features obtained by CPR, and c) relative correlations between pixel-wise samples. In this way, the bottleneck of regression performance caused by the limited reference data is mitigated.

Our main contributions are in pioneering application of CL in mapping continuous forest variables using EO data, developing a new CL loss enabling production of pixel-level forest maps, and proposing a new semisupervised hybrid framework, that effectively explores trainign EO datasets. 
Finally, we demonstrate the superiority of proposed models compared to traditional approaches over a boreal forest site. To the best of our knowledge, it is the first communication on successful application of CL in wall-to-wall forest mapping using SAR and optical satellite data.

This letter is organized as follows. Firstly we introduce CtRL loss, formulate suitable similarity function, and introduce hybrid semisupervised framework in Section II. Then we describe study site, used EO and reference datasets in Section III. Experimental results on applying both developed models and more traditional approaches and their analysis are gathered in section IV, and the paper is concluded in Section V. 

\section{Methodology}
\label{sec:method}
\subsection{Contrastive Regression Loss}
Given a set of random samples $\mathbb{Z}=\{z_1,\cdots, z_n\}$ that includes one positive sample $z_i$ drawn from conditional distribution $p(z \mid x)$ and $N-1$ negative samples drawn from proposal distribution $p(z)$, the probability of correctly classifying the positive sample $z_i$ from $p(z\mid x)$ rather than $p(z)$ is
\begin{equation}
p(d=i \mid \mathbb{Z}, x) =\frac{\frac{p\left(z_i \mid x\right)}{p\left(z_i\right)}}{\sum_{j=1}^N \frac{p\left(z_j \mid x\right)}{p\left(z_j\right)}} \simeq \frac{f_i\left(z_i, x\right)}{\sum_j f_j\left(z_j, x\right)},
\label{eq:dr}
\end{equation}
where x is the input, $[d=i]$ indicates $z_i$ is the positive sample. 
In InfoNCE, $p\left(z \mid x\right)/p\left(z\right)$ is called "density ratio" and assumed can be modeled by a log-bilinear model as $f\left(z, x\right) \propto {p\left(z \mid x\right)}/{p\left(z\right)}$ \cite{oord2018representation}. 

To extend CL to regression, given two different positive samples $i$ and $k$, we assume the ratio between distributions of $p\left(d=i \mid \mathbb{Z}, x\right)$ and $p\left(d=k \mid \mathbb{Z}, x\right)$ is proportional to the similarity between their label distributions $p(h_i)$ and $p(h_k)$:
\begin{equation}
\frac{p\left(d=i \mid \mathbb{Z}, x\right)}{p\left(d=k \mid \mathbb{Z}, x\right)}\simeq\frac{f_i\left(z_i, x\right)}{f_k\left(z_k, x\right)} \simeq \lambda \cdot \mathrm{S}\left[p\left(h_i\right) ; p\left(h_k\right)\right],
\label{eq:assum}
\end{equation}
where $\lambda$ is the scale factor and $\mathrm{S}(\cdot)$ is the similarity function. 
Substitute Eq. \ref{eq:assum} into Eq. \ref{eq:dr}, and take all $N-1$ negative samples into account, Eq. \ref{eq:dr} can be approximately converted into an average of $N-1$ similar expressions like
\begin{equation}
p\left(d=i \mid \mathbb{Z}, x\right)=\frac{\frac{\lambda}{N-1} \sum_k \mathrm{S}_{i,k} \cdot f_k\left(z_k, x\right)}{\sum_j f_j\left(z_j, x\right)},
\label{eq:prob}
\end{equation}
where $\mathrm{S}_{i,k}$ denotes the similarity function $\mathrm{S}\left[p\left(h_i\right) ; p\left(h_k\right)\right]$ for simplicity.
Note that $k$ and $j$ in fact indicate the same set of negative samples in the batch (excluding sample $i$).

Following \cite{chen2020simple}, $f(z, x)$ is defined as $\exp\left(\mathrm{C}_{i,k}/\tau\right)$, where $\mathrm{C}_{i,k} = CosSim(z_i, z_k)$ denotes the cosine similarity between normalized embedding features, and $\tau$ is the ``temperature" coefficient to control the sensitivity to negative samples. Based on negative log loss (NNL), the final form of CtRL derived from Eq. \ref{eq:prob} after removing the constant term is 
\begin{equation}
\mathcal{L_\mathrm{CtRL}} =-\log \frac{\sum_k \mathrm{relu}(\mathrm{S}_{i,k}) \cdot \exp\left(\mathrm{C}_{i,k}/\tau\right)}{\sum_j |\mathrm{S}_{i,j}|\cdot \exp\left(\mathrm{C}_{i,j}/\tau\right)},
\label{eq:ctrl}
\end{equation}
where $\mathrm{relu}$ operation is to discard negative similarities and avoid \textit{NAN} at the beginning of the optimization, $|\cdot|$ denotes the absolute operation to normalize $\mathcal{L_\mathrm{CtRL}}$.
Similar to InfoNCE in classification tasks, CtRL has been proved its ability to pull together closer samples and push away farther samples on embedding hypersphere, guided by their regression labels \cite{wang2022contrastive}.

As for the choice of similarity function $\mathrm{S}_{i,k}$, we introduce the negative log Kullback-Leibler (KL) divergence to model the asymmetry discrepancy between two label distributions. Assuming Gaussianity in forest mapping case, we denote the label distribution for $i$ and $k$ sample as $h_i \sim \mathcal{N}(\mu_i, \sigma_i)$ and $h_k \sim \mathcal{N}(\mu_k, \sigma_k)$, then $\mathrm{S}_{i,k}$ is calculated as 
\begin{equation}
\begin{aligned}
\mathrm{S}_{i,k} &= -\log \mathrm{KL}\left(\mathcal{N}\left(\mu_i, \sigma_i^2\right)|| \mathcal{N}\left(\mu_k, \sigma_k^2\right)\right)\\
&= -\log \frac{\left(\mu_i-\mu_k\right)^2}{2 \sigma^2},
\end{aligned}
\end{equation}
where $\mu_i$ and $\mu_k$ are the average forest height for $i$ and $k$ in our case, and $\sigma_i=\sigma_k=\sigma$ is the aggregated measurement error of airborne laser scanning (ALS) point cloud.

\subsection{Composition of Pixel-wise Negative Samples}
It is recognized that negative samples play a key role in avoiding the model collapsing \cite{chen2020simple, he2020momentum}. 
However recent works in remote sensing mainly focus on instance-level composition of positive/negative samples, which is not suitable for wall-to-wall regression tasks.
Here we propose a composition strategy of pixel-wise negative samples.

Similar to DL based semantic segmentation, wall-to-wall forest mapping is usually based on training over image patches \cite{scepanovic2021}.
Each pixel within the patch is an individual measurement sample.
Provided the batch size is $N_b$ and the size of image patch is $H\times W$, there are $N_b\cdot H\cdot W$ measurements within each training batch. The computation of CtRL for all the measurements easily becomes too large for GPU memory. 
In order to decrease the data scale, we randomly sample $N$ so-called anchor points across all the $N_b$ image patches.

\begin{figure}[htb]
\centering
\includegraphics[width=0.7\linewidth]{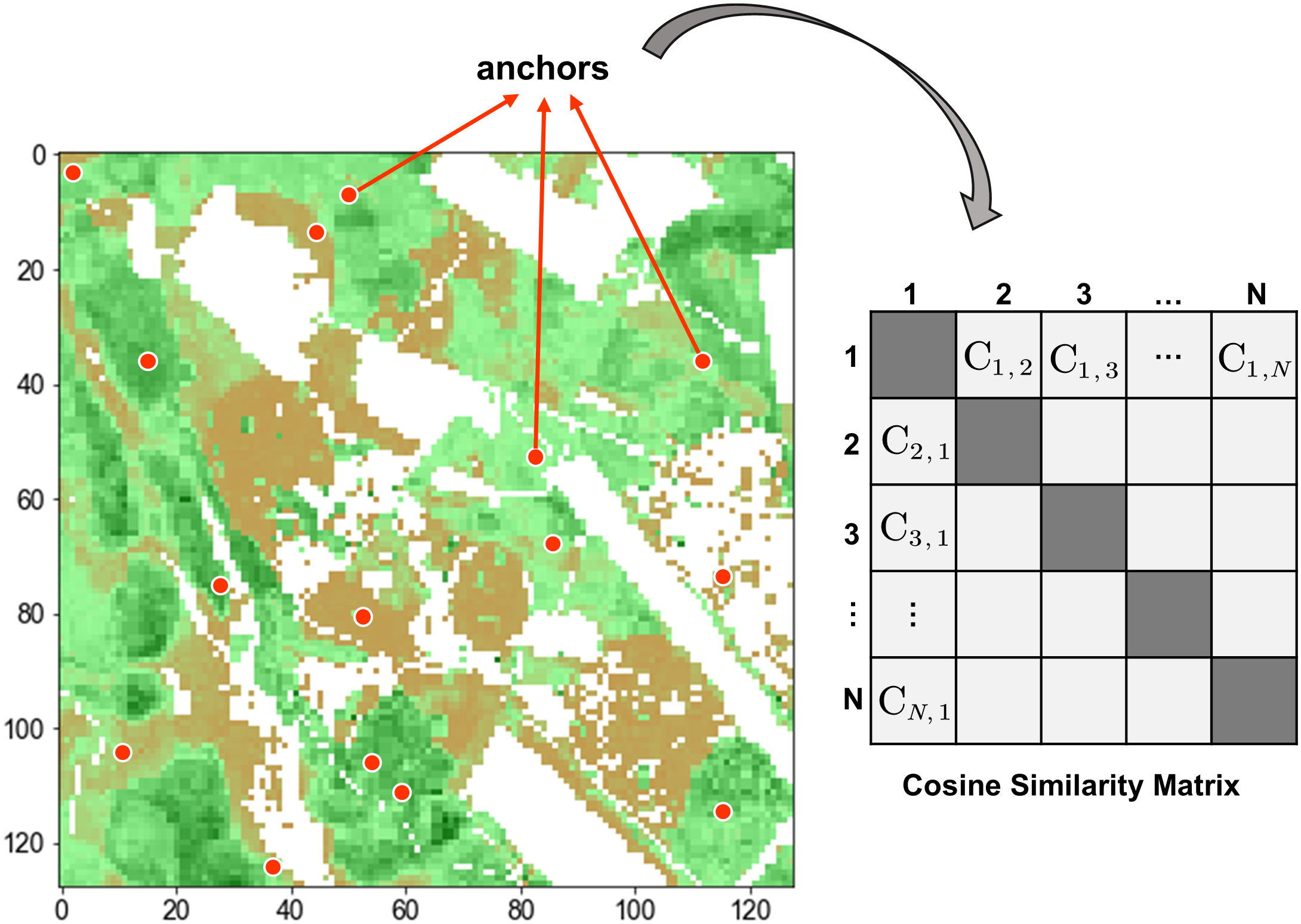}
\caption{An example of $N$ anchors randomly sampled from the image patch, and its cosine similarity matrix $\mathrm{C}_{i,k}$. 
Diagonal grids are excluded.
}
\label{fig:anchors}
\end{figure}
Then the cosine similarity can be computed in a parallel manner with a $N\times N$ matrix. Considering Eq. (\ref{eq:ctrl}) only takes negative samples into computation, the positive samples are neglected in this matrix. For each anchor, all the other anchors in the same batch are its corresponding negative samples. We will further analyze the impact of different amounts of anchors on the regression performance in Section \ref{sec:results}.

\subsection{Hybrid Semisupervised Framework}

\begin{figure*}[htb]
\centering
\includegraphics[width=0.9\textwidth]{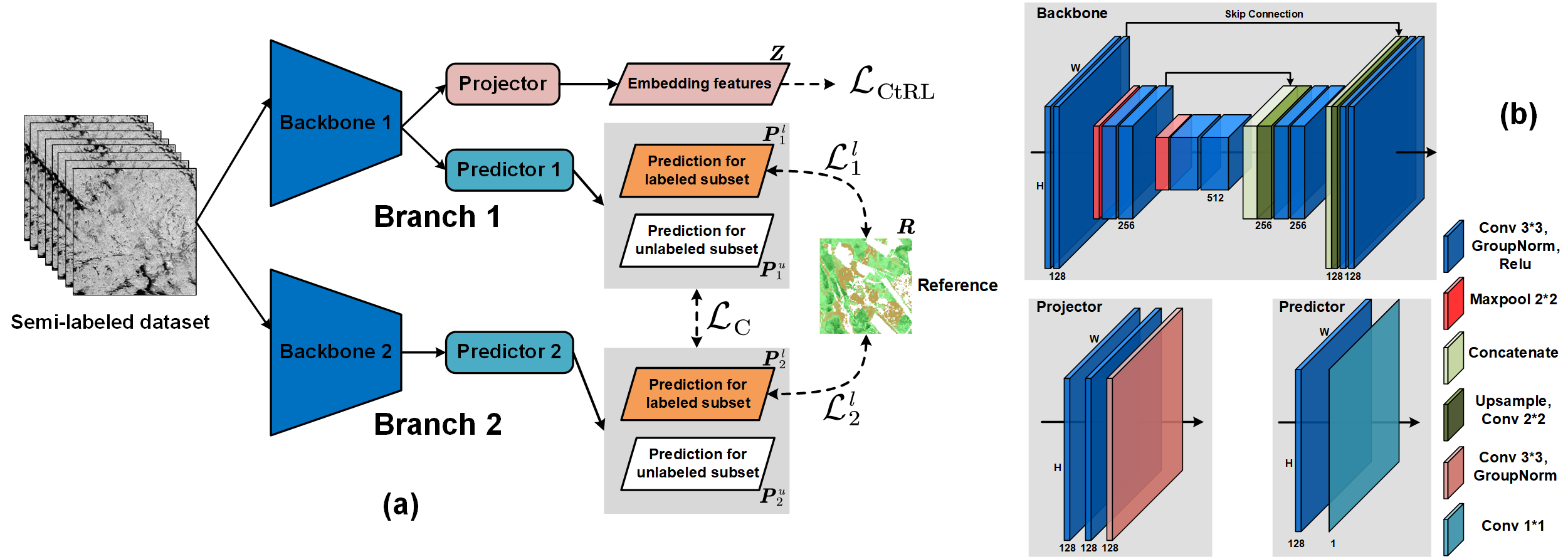}
\caption{The overall structure of the proposed semisupervised forest height mapping framework. (a) is the overall flowchart, (b) illustrates the detail structure of the components. The colored layers denote the feature maps after corresponding operations.}
\label{fig:model}
\end{figure*}

To embed CtRL into the regression model, a new projector head is added to a vanilla UNet as shown in \textit{Branch 1}, Figure \ref{fig:model}.
A three-layer vanilla UNet is used as the backbone network \cite{ronneberger2015u,ge2022improved}.
The basic convolutional unit of vanilla UNet commonly consists of one \textit{Conv2D} layer, followed by one \textit{BatchNorm} and \textit{Relu} activation function. We modify it by replacing \textit{BatchNorm} with \textit{GroupNorm}, as Group Normalization is more robust to smaller batch size \cite{wu2018group}. The number of norm groups is 4 in our case.
The convolutional kernel size is $3*3$. The numbers of kernels in each \textit{Conv2D} layer are denoted with footnotes in Figure \ref{fig:model}.b. 
As to the upsampling operation, we apply the deconvolution instead of bilinear interpolation. Its stride is set as $2$ to avoid the tessellation effect. The output feature map size of backbone is $H\times W\times 128$.

The representative features learned by the backbone is then fed into two heads including a projector and a predictor. The projector consists of three layers: the first two are the above mentioned convolutional units, the third layer includes one \textit{Conv2D} layer and one \textit{GroupNorm} but without any activation function. The output tensor $\boldsymbol{Z}$ of the projector has a shape of $H\times W\times 128$. By sampling feature vectors $\boldsymbol{z}_i \in \mathbb{R}^{1\times 1\times128 }$ from $\boldsymbol{Z}$, CtRL is calculated 
according to Eq. \ref{eq:ctrl}. The predictor simply consists of one convolutional unit and one \textit{Conv2D} layer with kernel size as $1*1$. It projects the upstream feature maps into pixel-wise predictions which has only 1 channel. 

Considering CtRL is a supervised loss where reference is involved in the computation, it is worth exploring its compatibility with a semisupervised framework like CPR. 
So a new branch, \textit{Branch 2}, is grafted to the model. Different from \textit{Branch 1}, only one predictor head exists in it.
For both branches, the predictions $\boldsymbol{P}_1$ $\boldsymbol{P}_2$ are composed of two parts: predictions with/without reference, which are represented as $\boldsymbol{P}_1 = \boldsymbol{P}^l_1 \cup \boldsymbol{P}^u_1$ and $\boldsymbol{P}_2 = \boldsymbol{P}^l_2 \cup \boldsymbol{P}^u_2$.
Let $\mathrm{M}(\cdot)$ denote the pixel-wise mean squared error (MSE), the CPR loss $\mathcal{L}_\mathrm{CPR}$ can be defined as
\begin{equation}
\begin{aligned}
\mathcal{L}_\mathrm{CPR} &= \mathcal{L}^l_1 + \mathcal{L}^l_2 + \lambda_c\mathcal{L}_\mathrm{C},\\
\mathcal{L}^l_1, \mathcal{L}^l_2, \mathcal{L}_\mathrm{C} &= \mathrm{M}(\boldsymbol{P}^l_1, \boldsymbol{R}), \mathrm{M}(\boldsymbol{P}^l_2, \boldsymbol{R}), \mathrm{M}(\boldsymbol{P}_1, \boldsymbol{P}_2),
\end{aligned}
\end{equation}
where $\boldsymbol{R}$ denotes the reference of the labeled training subsets, $\lambda_c$ is a balancing weight that is simply set as 1 in our case. For more details of CPR, one can refer to our previous work \cite{ge2022improved}. 

Finally, a hybrid semisupervised loss is proposed as a combination of CPR and CtRL losses:
\begin{equation}
\mathcal{L}_\mathrm{Semi} = \mathcal{L}_\mathrm{CPR} + \lambda_{ctrl}\mathcal{L}_\mathrm{CtRL} + \lambda_w \frac{1}{n_w} \sum_{j=1}^{n_w}\left(w_j\right)^2.
\label{eq:loss}
\end{equation}
The trade-off weight $\lambda_{ctrl}$ should guarantee $\mathcal{L}_\mathrm{CtRL}$ and $\mathcal{L}^l_1$ on the same order of magnitude. We set it as 60 according to cross-validation. The last term is a weight decay to mitigate the overfitting, where $n_w$ is the total number of weights. After training, \textit{Branch 2} is extracted as the final model for the regression task. 
The proposed framework is named as CPrUNet+CtRL.
In the following sections, we will validate its effectiveness in better mining the limited reference information.

\section{Study Site, EO and Reference Data}

\begin{figure}[htb]
\centering
\includegraphics[width=\linewidth]{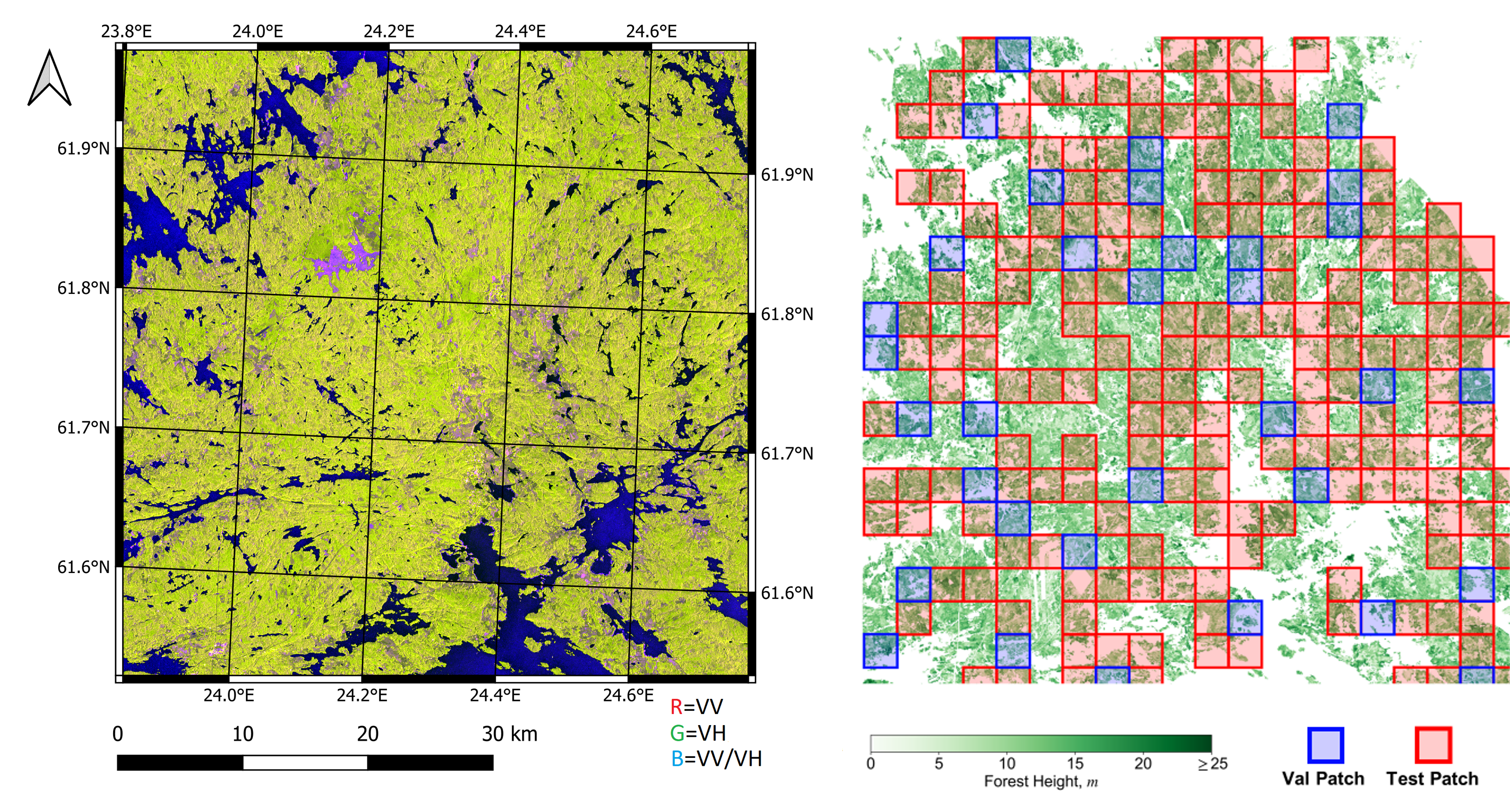}
\caption{Sentinel-1 RGB composite over study site and schematic allocation of training and accuracy assessment image patches.}
\label{fig:data}
\end{figure}

Our study area of \SI{50}{km}$\times$\SI{50}{km} size (shown in Figure \ref{fig:data}) is located in the Tampere region in the central part of Finland. Here, boreal forest is represented by a mixture of Norway spruce, Scots pine and birch, with average growing stock of \SI{170}{m^3/ha}.
In our experiments, we used a combination of Copernicus SAR and optical data that were found more useful than SAR or optical data separately in our earlier and other similar studies \cite{ge2022improved}. 
SAR data were represented by a time series of 27 dual-pol (VV, VH) Sentinel-1 ground range detected (GRD) images covering the whole year of 2015. The images were orthorectified and radiometrically terrain corrected using Europe Space Agency (ESA) SNAP software and local digital elevation model from National land Survey of Finland. The final preprocessed images were in gamma-naught format with pixel spacing 20$\times$\SI{20}{m^2}.
Further, a Sentinel-2 Level-2A optical image acquired in July 2015 was used in the study. Four \SI{10}{m} spatial resolution bands were included as image features in various examined regression models.

Reference data were represented by airborne laser scanning data acquired by National Land Survey of Finland during summer of 2015. For each 20$\times$\SI{20}{m^2} spatial area, mean relative height over ground was calculated for all forest classified cloud points.
The forest stand mask was additionally applied with the purpose of removing non-forest areas such as urban settlements or water bodies, and also used for computing stand-level estimates. 

The whole study area was divided into image patches shown in Figure \ref{fig:data}.c, each patch has a size of \SI{128}{}$\times$\SI{128}{px}. 
For model training and accuracy assessment, we used only patches with at least 20\% forest cover, resulting in 340 non-overlapping patches, further split  into training, testing and validation sets as follows: 50\% randomly sampled as testing set, 10\% for validation set, and the rest were used for training.
Data augmentation included shifting and rotation, resulting in 716 training, 34 validation, and 170 testing image patches.

\subsection{Baselines and Accuracy Metrics}
We compared the proposed semisupervised regression framework with three baseline approaches: Multiple Linear Regression (MLR), Random Forest (RF) and Light Gradient Boosting Machine (LightGBM). Within MLR, a principal component analysis was applied as additional preprocessing step to reduce the amountof input image features. Feature extraction was not applied with RF and LightGBM, as these methods have a built-in feature selection mechanism.
As considered baseline methods operate on pixel-level, the training/test/validation subsets were constructed by reshaping all pixels from corresponding image patches.
In addition, vanilla UNet and CPR-strategy based UNet (CPrUNet) \cite{ge2022improved} were also included in the ablation study to evaluate the effectiveness of the proposed framework.

The prediction accuracy is evaluated using root mean square error (RMSE), relative RMSE (rRMSE), mean absolute error (MAE), coefficient of determination (R$^2$) and index of agreement (IOA). Evaluation is done both on pixel level and forest stand level. Stand-level metrics are calculated after averaging all pixel-level predictions within each forest stand.

\section{Experimental Results and Discussion}
\label{sec:results}

\subsection{Experimental Settings}
Experiments were run on a 64-bit Win10 desktop with \SI{32}{GB} RAM accelerated by a NVIDIA GTX3060 GPU. The proposed model was built by PyTorch 1.11 and Python 3.8. 
After preliminary testing, the quantity of anchors was set to 1000. During the training, \textit{Adam} was chosen as the optimizer and \textit{OneCycleLR} as the learning rate scheduler, the maximum learning rate was set as $10^{-2}$. We trained the model for 100 epochs, the weight decay factor was set to $10^{-4}$.  Checkpoints were saved and updated according to the validation loss, with the best checkpoint used in the testing stage.

\subsection{Classification Performance Analysis}
To demonstrate the utility of proposed semisupervised regression framework, we start with an ablation study where CtRL loss and CPR strategy were tested separately. The quantitative experimental results in forest height mapping are gathered in Table \ref{tab:results}, with the best accuracy figures in bold.
\begin{table}[]
\centering
\caption{Prediction performance compared to ablation benchmarks and conventional methods.}
\label{tab:results}
\renewcommand\arraystretch{1.2}
\resizebox{\columnwidth}{!}{%
\begin{tabular}{clccccccc}
\toprule
 &  & \textit{CtRL} & \textit{CPR} & \textbf{\begin{tabular}[c]{@{}c@{}}RMSE \\ (m)\end{tabular}} & \textbf{\begin{tabular}[c]{@{}c@{}}rRMSE \\ (\%)\end{tabular}} & \textbf{\begin{tabular}[c]{@{}c@{}}MAE \\ (m)\end{tabular}} & \textbf{R$^2$} & \textbf{\begin{tabular}[c]{@{}c@{}}IOA \\ (\%)\end{tabular}} \\ \midrule
\multirow{7}{*}{\textbf{Pixel-level}} & \textbf{MLR} &  &  & 3.57 & 31.96 & 2.79 & 0.38 & 74.25 \\
 & \textbf{RF} &  &  & 3.40 & 30.42 & 2.66 & 0.44 & 76.20 \\
 & \textbf{LightGBM} &  &  & 3.35 & 30.00 & 2.60 & 0.46 & 78.03 \\
 & \textbf{UNet} &  &  & 2.79 & 25.00 & 2.06 & 0.62 & 87.50 \\
 & \textbf{CPrUNet} &  & $\checkmark$ & 2.74 & 24.51 & 2.01 & 0.64 & 88.24 \\
 & \textbf{UNet+CtRL} & $\checkmark$ &  & 2.72 & 24.37 & 1.98 & 0.64 & 88.74 \\
 & \textbf{CPrUNet+CtRL} & $\checkmark$ & $\checkmark$ & \textbf{2.66} & \textbf{23.86} & \textbf{1.94} & \textbf{0.66} & \textbf{89.01} \\ \midrule
\multirow{7}{*}{\textbf{Stand-level}} & \textbf{MLR} &  &  & 2.41 & 21.58 & 1.88 & 0.50 & 79.83 \\
 & \textbf{RF} &  &  & 2.28 & 20.41 & 1.79 & 0.55 & 81.20 \\
 & \textbf{LightGBM} &  &  & 2.21 & 19.76 & 1.72 & 0.58 & 83.35 \\
 & \textbf{UNet} &  &  & 1.79 & 16.02 & 1.31 & 0.73 & 91.39 \\
 & \textbf{CPrUNet} &  & $\checkmark$ & 1.73 & 15.53 & 1.26 & 0.74 & 92.12 \\
 & \textbf{UNet+CtRL} & $\checkmark$ &  & 1.74 & 15.63 & 1.26 & 0.74 & 92.28 \\
 & \textbf{CPrUNet+CtRL} & $\checkmark$ & $\checkmark$ & \textbf{1.68} & \textbf{15.07} & \textbf{1.22} & \textbf{0.76} & \textbf{92.69} \\ \bottomrule
\end{tabular}%
}
\end{table}

Vanilla UNet model provided  rRMSEs of 25.0\% and 16.0\% at pixel- and stand-level, respectively.
After introducing the CPR strategy, rRMSEs of CPrUNet improved to 24.5\% and 15.5\%, explained by its ability in exploiting features from unlabeled samples. However, straightforward integration of CtRL loss into vanilla UNet (UNet+CtRL) provided similar rRMSEs of 24.37\% and 15.63\%. It suggests an important role of CtRL in measuring distance between labeled anchors, improving the prediction performance compared to vanilla UNet. Importantly, the hybrid CPrUNet+CtRL model combines advantages of both CPR and CtRL to deliver superior prediction accuracy, with rRMSEs of 23.86\% and 15.07\%. It indicates the suggested hybrid framework can learn more comprehensive feature set from the available data.

\begin{figure}[htb]
\centering
\includegraphics[width=0.8\linewidth]{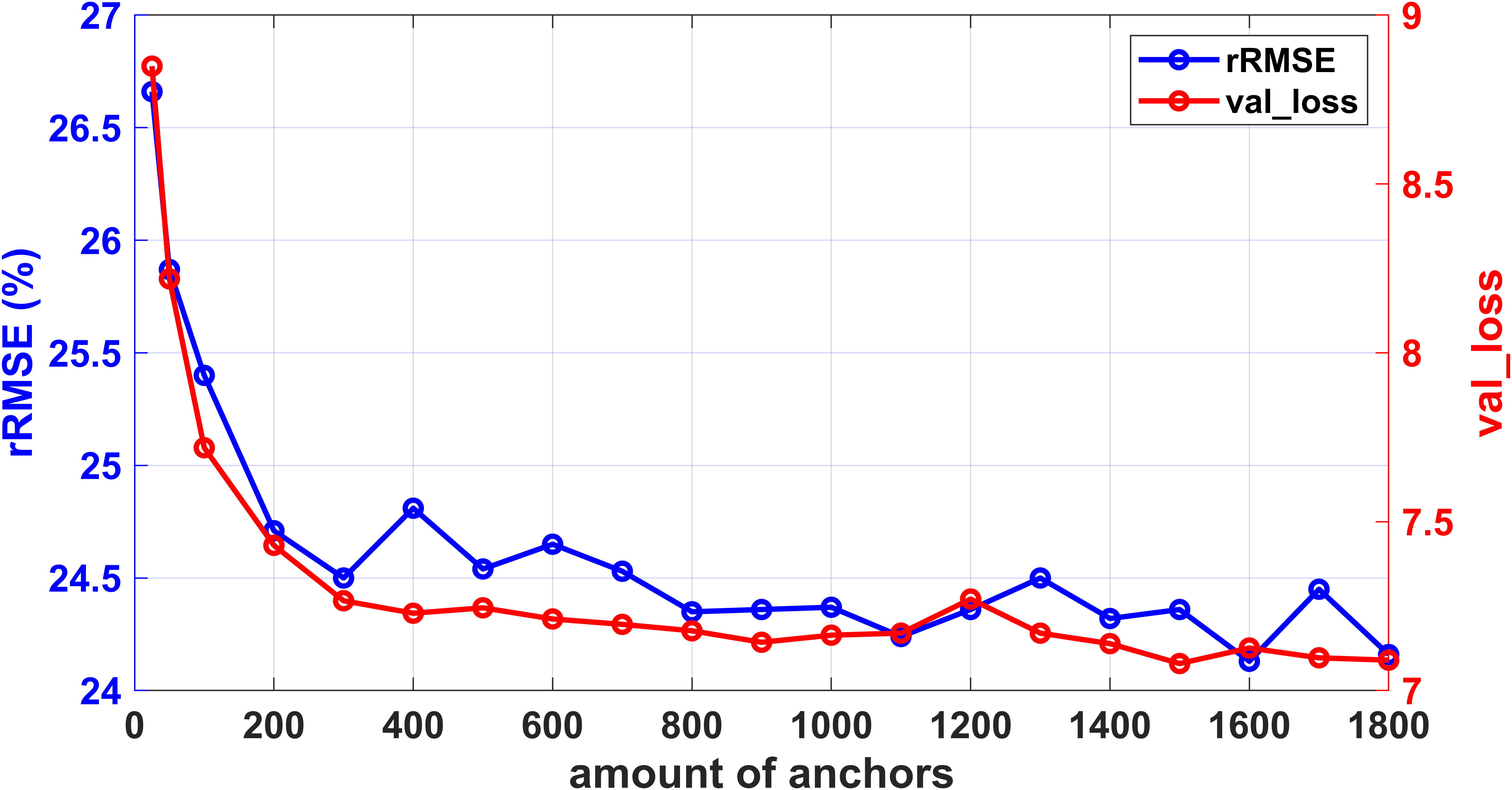}
\caption{Dependence of pixel-level accuracy metrics (rRMSE and validation loss) on the quantity of anchors.}
\label{fig:anchors_curve}
\end{figure}

Since the computation of CtRL loss relies on randomly sampled anchors, we evaluated the impact of amount of anchors on prediction accuracy in more detail.
Figure \ref{fig:anchors_curve} shows dependence of pixel-level prediction statistics on the amount of anchors using rRMSE and validation loss as representative measures, for UNet+Ctrl model. 
As the amount of anchors increased, both measures decreased, saturating as the number of anchors reached 1000.
Although sampling more anchors seems generally better for the accurate calculation of CtRL loss, their amount had to be restricted to prevent the parallel matrix computation overflowing the limited GPU memory. In experiments performed within the ablation study, 1000 anchors were sampled for each batch as an outcome of trade-off analysis.

Further, we compared the developed methodologies with MLR, RF and LightGBM (also gathered in Table \ref{tab:results}). 
As these methods do not consider spatial context, the prediction accuracies were understandably poorer with all rRMSEs larger than 30.0\% at pixel-level and 19.0\% at stand-level. Among those more conventional methods, LightGBM was the best prediction approach with the accuracy gain of 1.96\% units compared to MLR. 
Further, introducing spatial context with the help of UNet improved prediction accuracy by 5\% RMSE units compared to LightGBM. Finally, proposed CPrUNet+CtRL additionally reduced rRMSE by 1.14\% units.
Our best stand-level predictions reach RMSE of \SI{1.68}{m} and R$^2$ of 0.76, superior to earlier reported studies in boreal zone \cite{astola2019,astola2021}. 

\begin{figure}[htb]
\centering
\includegraphics[width=\linewidth]{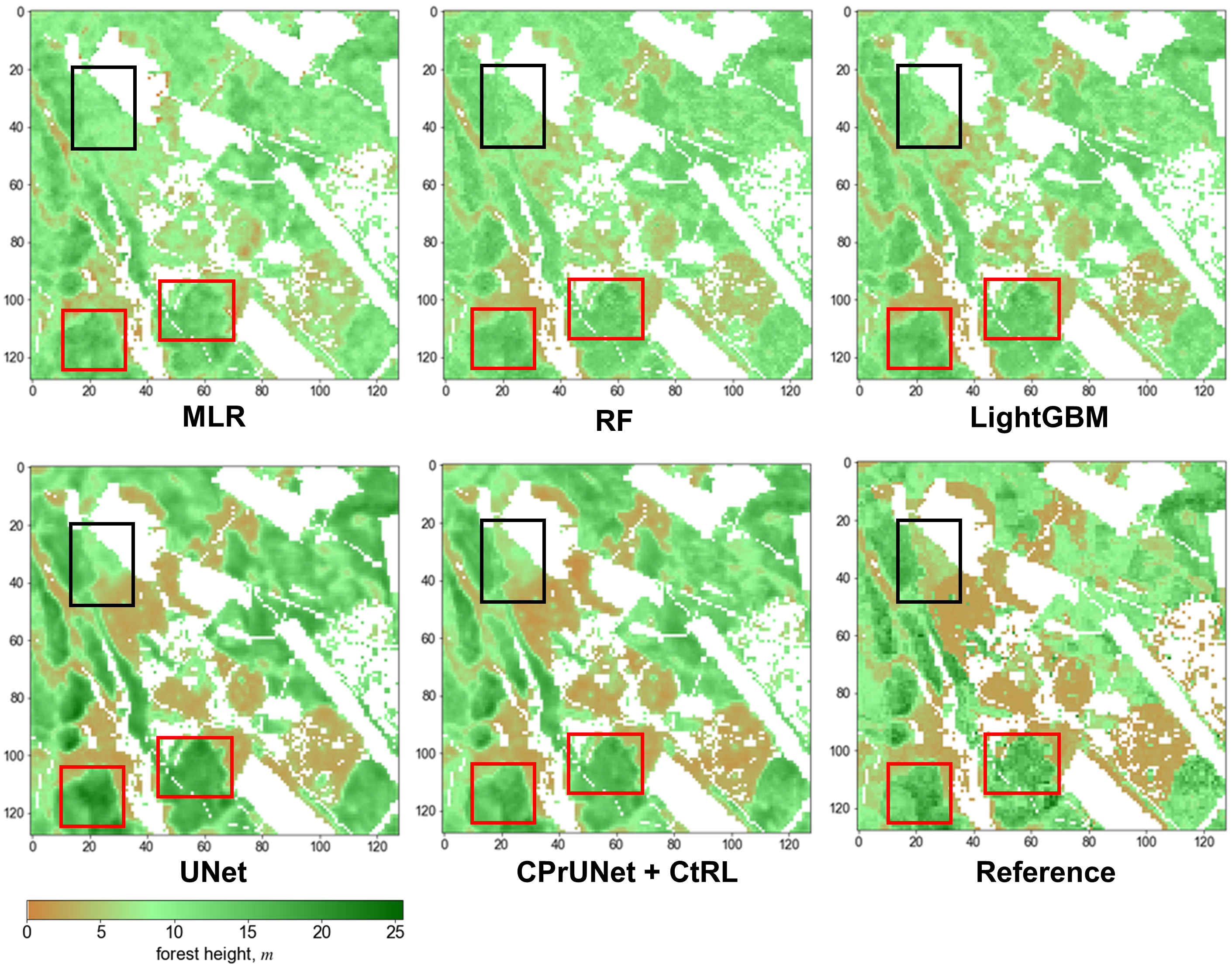}
\caption{Examples of forest height maps produced by studied regression methods. Closely examined areas are highlighted with polygons.}
\label{fig:results}
\end{figure}

\begin{figure}[htb]
\centering
\includegraphics[width=\linewidth]{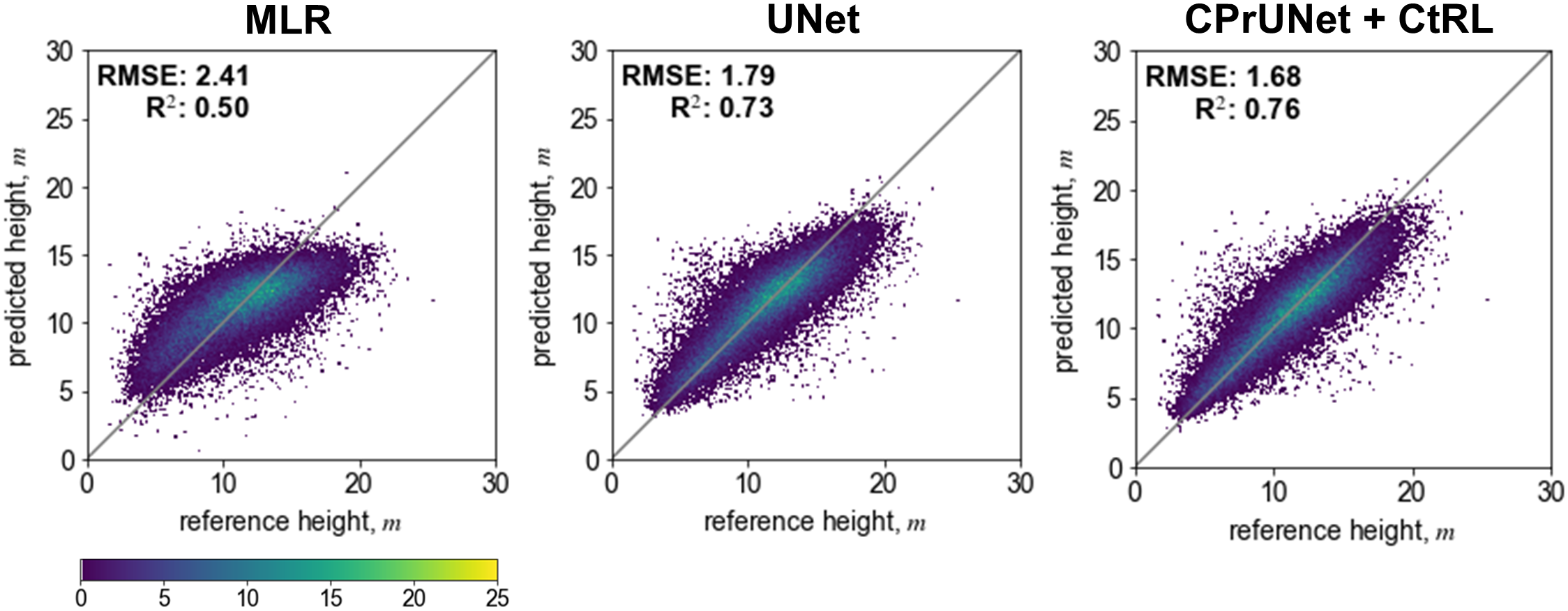}
\caption{Scatterplots illustrating the tree height prediction performance on stand-level for selected regression models.}
\label{fig:scatterplots}
\end{figure}

Example of prediction results over a single image patch is shown in Figure \ref{fig:results}. As highlighted in red polygons, CPrUNet+CtRL better reveals details of taller forests compared to vanilla UNet. Similarly, proposed model provided better predictions of shorter trees (black polygon in Fig. 5), better highlighting differences between neighbouring forest compartments and revealing logged areas.
Representative scatterplots illustrating prediction accuracy of MLR, UNet and developed framework are shown in Figure \ref{fig:scatterplots}.
Saturation of forest height predictions is clearly visible for MLR, and less apparent for vanilla UNet, mostly remaining for trees higher than \SI{15}{m}. In contrast, CPrUNet+CtRL approach overcomes this drawback indicating that particularly heights of taller trees can be better predicted.

\section{Conclusion}
In this study, a contrastive regression approach is introduced into EO-based forest inventory, and a hybrid semisupervised framework is developed enabling more accurate preditions of forest attributes compared to earlier studied DL models or traditional regression approaches. The framework is demonstrated over a Finnish boreal site for predicting forest tree height using ESA Sentinel-1 and Sentinel-2 data and improving prediction accuracy. We expect the developed framework can be used to model relationships between EO data and other forest attributes, and be applicable also in other forest biomes.

\bibliographystyle{IEEEtran}
\bibliography{ctrl.bib}

\newpage

\end{document}